\documentclass[aps,pra,twocolumn,10pt,showpacs]{revtex4-1}
\usepackage{amsmath}
\usepackage{graphicx}
\usepackage{times}
\usepackage{amssymb}
\usepackage{hyperref}
\usepackage{textcomp}
\usepackage{xcolor}
\usepackage{enumerate}
\usepackage[final]{pdfpages}

\begin{document}

\title{Operational Gaussian Schmidt-Number Witnesses}
\author{F. Shahandeh}
\email{Electronic address: farid.shahandeh@uni-rostock.de}
\author{J. Sperling}
\author{W. Vogel}
\affiliation{Arbeitsgruppe Theoretische Quantenoptik, Institut f\"ur Physik, Universit\"at Rostock, D-18055 Rostock, Germany}
\date{\today} 

\begin{abstract}
The general class of Gaussian Schmidt-number witness operators for bipartite systems is studied.
It is shown that any member of this class is reducible to a convex combination of two types of Gaussian operators using local operations and classical communications.
This gives rise to a simple operational method, which is solely based on measurable covariance matrices of quantum states.
Our method bridges the gap between theory and experiment of entanglement quantification.
In particular, we certify lower bounds of the Schmidt number of squeezed thermal and phase-randomized squeezed vacuum states, as examples of Gaussian and non-Gaussian quantum states, respectively.
\end{abstract}


\maketitle

\section{Introduction}
Entanglement detection and quantification are of great relevance for quantum information and quantum communication science~\cite{Bennett1,DiVincenzo,Nielsen}.
Two problems of significant importance then arise.
First, one is interested to find out whether a quantum state is separable or entangled.
Second, if the state is entangled, one needs to quantify the amount of entanglement.
A large amount of work is devoted to the first problem, 
dealing with the separability criteria~\cite{Peres,Horodecki1,Rudolph,Chen,Shchukin,Guhne1}.
The answer to the second question requires proper entanglement measures~\cite{Terhal,Plenio}, which is rather complicated in general.

A possibility to answer this problems is the use of the Schmidt number (SN)~\cite{Sanpera,Terhal,Bruss}. It generalizes the Schmidt rank~\cite{Nielsen} to bipartite mixed states, based on a convex roof construction~\cite{Terhal,Uhlmann1,Uhlmann2}.
Therefore, the SN is closely related to the quantum superposition principle. It gives the minimal number of superpositions needed to construct the state under study.  
In other words, it is directly connected to the number of global superpositions of two entangled subsystems.
It has been shown that the SN fulfills the axioms of entanglement measures~\cite{Vedral,Hulpke,Horodecki2,Guhne2}.
Moreover, it does not increase under all separable operations,
so that it is a universal entanglement measure ~\cite{Sperling3}.
Note that the convex sets of states having a SN less than or equal to $r$, $\hat{\varrho} \in {\mathcal S}_r$, 
have a nested structure of the form ${\mathcal S}_1 \subset {\mathcal S}_2 \dots \subset{\mathcal S}_r \dots \subset {\mathcal S}_\infty$.
Thus, to verify to which set a state belongs, one can apply the method of SN witnesses~\cite{Sperling1}. For practical applications, however, this requires specifying the accessible observables.

In practice, continuous variable (CV) entangled states are often composed of squeezed states.
Hence, the latter are a key resource for CV quantum information processing and communication~\cite{Wang}. 
They are also fundamental in quantum teleportation protocols~\cite{Furusawa, Bennett2}.
Measurements are usually based on balanced homodyne detection, which easily yields the covariance matrix (CM).  
The squeezed states and the resulting CV entangled states are usually Gaussian and hence they are completely characterized by their CM~\cite{Braunstein,Adesso}.
The separability problem of bipartite Gaussian states was completely solved~\cite{Duan,Simon2}.
However, even for this class of states, an experimentally accessible entanglement quantification is still missing. 
In fact, CV entanglement measures for broad experimental applications are presently unknown~\cite{Stobinska}.

In the present contribution, we fully characterize the bipartite Gaussian SN witness operators.
We show that, up to second order moments, there exist two generating classes of SN witnesses so that any Gaussian witness operator can be expressed as a convex combination of them.
This relates SN witnessing to the measurable CM of the quantum state under study, which leads to a simple operational method of entanglement quantification.
For Gaussian state this method is complete. For non-Gaussian states it still yields useful lower boundaries, which certify a definite amount of entanglement.
Rather than searching for the best witness for any quantum state, we aim at extracting maximal information from the CM, which is usually recorded in CV experiments.
We apply our approach to the entanglement quantification of examples of Gaussian and non-Gaussian mixed quantum states.

The paper is organized as follows.
In Sec.~\ref{GSNw} we review the SN witnesses and their optimization.
Then, we introduce the general form of Gaussian SN witnesses and solve the set of SN eigenvalue equations.
The process of SN witnessing using the measurable CM is considered in Sec.~\ref{Exam}.
Some examples are discussed in this section.
A summary and conclusions are given in Sec.~\ref{Summ}.

\section{Gaussian SN witnesses}
\label{GSNw}

In the following, we briefly review the SN witnesses and their optimization problem (see Ref.~\cite{Sperling1} and references therein).
In particular, we will study the class of Gaussian SN witnesses.
This yields an efficient method to count global quantum superpositions.

\subsection{Optimized SN witnesses}
A SN-$r$ witness $\hat{\mathcal W}$ is a Hermitian operator such that $\langle \hat {\mathcal W}\rangle := {\text{Tr}}(\hat \varrho \hat {\mathcal W}) \geqslant 0$, with $\hat{\varrho} \in {\mathcal S}_r$, that is for all states having a SN less than or equal to $r$.
It is also required that $\langle \hat {\mathcal W}\rangle < 0$ for at least one state with a SN greater than $r$, $\hat \varrho \notin {\mathcal S}_r$.
Since any witness operator can be written as 
\begin{equation}
\label{witnessform}
  \hat{\mathcal W}=\hat{\mathcal L}{-}\lambda \hat{\mathbb I}, 
\end{equation}
with $\hat{\mathcal L}$ being a positive semidefinite operator, one can optimize any SN witness operator by choosing $\lambda$ to be the infimum of $\hat{\mathcal L}$ over all pure (and consequently mixed) quantum states.
That is, if ${\mathcal S}_r^{{\text{pure}}}$ denotes the set of pure states having a SN less or equal to $r$, we get
\begin{equation}
  \lambda = {g_r} := \inf \left\{ {\left\langle \chi  \right|\hat{\mathcal L}\left| \chi  \right\rangle : \left| \chi  \right\rangle \left\langle \chi  \right| \in {\mathcal S}_r^{{\text{pure}}}} \right\}.
\end{equation}
Hence, a quantum state $\hat{\varrho}$ has a SN greater than $r$, if and only if there exists a positive semidefinite Hermitian operator $\hat{\mathcal L}$ such that
\begin{equation}
  \langle \hat{\mathcal L} \rangle <g_r.
\end{equation}
Obviously, by this optimization, there exists no finer witness constructed from $\hat{\mathcal L}$ in the form of Eq.~\eqref{witnessform}.
In other words, the number of states $\hat{\varrho} \notin {\mathcal S}_r$ which can be detected by $\hat{\mathcal W}$ is maximal. 
Hereafter, we only speak of the \textit{test} operator $\hat{\mathcal L}$ rather than witness operator $\hat{\mathcal W}$, having in mind that they are related through Eq.~\eqref{witnessform}.

Following Ref.~\cite{Sperling1}, the optimization leads to the set of SN eigenvalue equations for $r \geqslant 1$:
\begin{align}
  \hat {\mathcal L}_{\vec \xi} | \vec{{\zeta _r}} \rangle  = g_{r}\mathbb{I}_{\vec \xi} | \vec{{\zeta _r}} \rangle,  \hfill \label{SNEE1}\\
  \hat {\mathcal L}_{\vec \zeta} | \vec{{\xi _r}} \rangle  = g_{r}\mathbb{I}_{\vec\zeta}  | \vec{{\xi _r}} \rangle,  \hfill \label{SNEE2}
\end{align}
in which each component of $| \vec{{\zeta _r}} \rangle  {=} ( {\begin{array}{*{20}{c}} {\left| {{\zeta _1}} \right\rangle }& \cdots &{\left| {{\zeta _r}} \right\rangle } \end{array}} )^T$ and $| \vec{{\xi _r}} \rangle  {=} ( {\begin{array}{*{20}{c}} {\left| {{\xi _1}} \right\rangle }& \cdots &{\left| {{\xi _r}} \right\rangle } \end{array}})^T$ belongs to the first and second Hilbert space, respectively.
Moreover, for $\hat {\mathcal A}\in\{ \hat {\mathcal L},\mathbb{I}\}$ and $i,j{=} 1,2,\ldots,r$, we have used the definitions
\begin{align}
	&\hat {\mathcal A}_{\vec \xi;i,j} ={\text{Tr}}_2({\hat {\mathcal A}\left[\mathbb I_1 \otimes \left| {{\xi_j }} \right\rangle \left\langle {{\xi_i}} \right|\right]} )\\
	&\hat {\mathcal A}_{\vec \zeta;i,j} = {\text{Tr}}_1({\hat {\mathcal A}\left[\left| {{\zeta_j }} \right\rangle \left\langle {{\zeta_i}} \right|\otimes\mathbb I_2\right]} ),
\end{align}
being the components of the block operators $\hat{\mathcal A}_{\vec \xi}$ and $\hat {\mathcal A}_{\vec \zeta}$, which act on $| \vec{{\zeta _r}} \rangle $ and $| \vec{{\xi _r}} \rangle $, respectively.
For any given SN $r$, the lowest common eigenvalue of Eqs.~\eqref{SNEE1} and~\eqref{SNEE2} is the value optimizing the witness operator $\hat{\mathcal W}$.
In the lowest order, $r=1$, Eqs.~(\ref{SNEE1}) and (\ref{SNEE2}) reduce to the separability eigenvalue equations as introduced in~\cite{Sperling2}.

\subsection{Bipartite Gaussian tests}

To obtain analytical solutions of Eqs.~\eqref{SNEE1} and~\eqref{SNEE2} is, in general, a sophisticated task.
However, it is possible to solve them for two maximally correlated cases and to provide the most general SN~test based on the CM.
We consider the general class of two-mode positive semidefinite Hermitian operators given in the form
\begin{equation} \label{QF}
  \hat {\mathcal L} = {{{\mathbf{\hat x}}}^T}\mathbf{\Omega} {\mathbf{\hat x}} + {{\mathbf{a}}^T}{\mathbf{\hat x}}+ C,
\end{equation}
where ${{{\mathbf{\hat x}}}^T} = ( {\begin{array}{*{20}{c}}{{{\hat q}_1}}&{{{\hat p}_1}}&{{{\hat q}_2}}&{{{\hat p}_2}} \end{array}} )$ define the position and conjugate momentum operators,
${{\mathbf{a}}^T} = ( {\begin{array}{*{20}{c}}\alpha_1 &\beta_1 &\alpha_2 & \beta_2 \end{array}} ) \in \mathbb{R}^4$ is an arbitrary vector,
and $\mathbf{\Omega}$ is a real Hermitian matrix:
\begin{equation}
  \mathbf\Omega  = \left( {\begin{array}{*{20}{c}}
  {\mathbf{\Omega}_1}&{\mathbf{\Omega}_c} \\ 
  {\mathbf{\Omega} _c^T}&{\mathbf{\Omega}_2} 
\end{array}} \right).
\end{equation}
It is well-known that for any local linear unitary transformation $\hat{U}_j$ ($j{=}1,2$), there exists a symplectic transformation $\mathbf{S}_j \in \text{Sp}(2,\mathbb R)$ with $\text{det}(\mathbf{S}_j) {=} 1$, so that ${{\hat U}_j}{( {\begin{array}{*{20}{c}}
  {{{\hat q}_j}}&{{{\hat p}_j}} 
\end{array}} )^T}\hat U_j^\dag  = {{\mathbf{S}}_j}{( {\begin{array}{*{20}{c}}
  {{{\hat q}_j}}&{{{\hat p}_j}} 
\end{array}} )^T}$~\cite{Simon1}.
Accordingly, the CM of any state $\hat{\varrho}$ can be written in the form
\begin{align}
\label{StForm}
  &\mathbf{V} = \left( {\begin{array}{*{20}{c}}
  {{\mathbf{V}_1}}&{{\mathbf{V}_c}} \\ 
  {\mathbf{V}_c^T}&{{\mathbf{V}_2}} 
\end{array}} \right),
\end{align}
where $\mathbf{V}= {\text{Tr}}( {\hat \varrho_{\rm s} {\mathbf{\hat x}}{{{\mathbf{\hat x}}}^T}})$,
such that ${\mathbf{V}_1} = v_1 {\mathbb{I}_{2 \times 2}}$, ${\mathbf{V}_2} = v_2 {\mathbb{I}_{2 \times 2}}$ and ${\mathbf{V}_c} = \text{diag}(v_{c,1},v_{c,2})$ form diagonal submatrices~\cite{Duan,Simon2,Pirandola}. Here,  $\hat{\varrho}_{\rm s}$ is the standard form of the density operator as used to form the standard form of the CM. It is obtained from $\hat{\varrho}$ under some appropriate local, unitary transformations. In this form $\left\langle {{{\hat q}_j}} \right\rangle  {=} \left\langle {{{\hat p}_j}} \right\rangle  {=} 0$ and $\left\langle {{{\hat q}_j}{{\hat p}_k}} \right\rangle {=} 0$ for $j,k{=}1,2$.
This leads to the following result: 
For any Gaussian test operator of the form~\eqref{QF} one can write
\begin{align} \label{ExL}
  {\text{Tr}}( {{{\hat \varrho }_{\rm s}}\hat {\mathcal L}} ) = {\omega _1}\left\langle {\left( {\hat q_1^2 + \hat p_1^2} \right)} \right\rangle  + {\omega _2}\left\langle {\left( {\hat q_2^2 + \hat p_2^2} \right)} \right\rangle \nonumber \\ + {\omega _{c;1}} \left\langle {{\hat q}_1}{{\hat q}_2} \right\rangle + {\omega_{c;2}} \left\langle {{\hat p}_1}{{\hat p}_2} \right\rangle,
\end{align}
with $\mathbf{\Omega}_1 {=} \omega_1 \mathbb I_{2 \times 2}$, $\mathbf{\Omega}_2 {=} \omega_2 \mathbb I_{2 \times 2}$, and $\mathbf{\Omega}_c {=} \text{diag} ( \omega_{c;1}, \omega_{c;2})/2$.

For the proof, let us fix the representation of the state to the standard form $\hat{\varrho}_{\rm s}$.
Obviously, the transformation from original to the standard form of the density operator, $\hat{\varrho} {\to} \hat{\varrho}_{\rm s}$, does not affect the separable or entangled structure of the state and thus the SN, because we only dealt with local operations.
Consider the expectation value ${\text{Tr}}( {{{\hat \varrho_{\rm s}}}\hat {\mathcal L}} )$, where $\hat {\mathcal L}$ is given in Eq.~\eqref{QF}.
The last term in $\hat {\mathcal L}$, regardless of the state, gives a constant $C$.
Then, without loss of generality, it can be set to be zero.
The linear term, as well as those corresponding to off-diagonal elements of $\mathbf{\Omega}_j$ ($j{=}1,2,c$), will vanish by using a local displacement operation and the standard form of the CM.
Therefore, we can choose these elements to be zero.

Now, the diagonal real matrices are reduced to
\begin{align}
 &\mathbf\Omega_j=\left( {\begin{array}{*{20}{c}}
  \omega_{j;1}&0 \\ 
  0&\omega_{j;2} 
\end{array}} \right), \quad j=1,2,c.
\end{align}
For $j=1,2$, the latter can be transformed into $\mathbf\Omega_j=\omega_j\mathbb{I}_{2 \times 2}$ with two appropriate local squeezing operations which is equivalent to apply the following (local) symplectic transformation to $\hat {\mathcal L}$:
\begin{align}
 &\mathbf{S}=\bigoplus^2_{j=1}\mathbf{S}_j,\\
 &\mathbf{S}_j = \left( {\begin{array}{*{20}{c}}
  s_j&0 \\ 
  0&1/s_j 
\end{array}} \right),\quad s_j=\left(\frac{\omega_{j;2}}{\omega_{j;1}}\right)^{\frac{1}{4}}.
\end{align}
Note that, hereafter, local orthogonal matrices $\mathbf{M}_1$ and $\mathbf{M}_2$ acting on the first and second modes, respectively, do not affect $\mathbf\Omega_1$ and $\mathbf\Omega_2$. At this point, the coupling matrix $\mathbf{\Omega}_c $ takes the form
\begin{align}
 \mathbf{\Omega}'_c{=}\mathbf{S}^{T}_1\mathbf{\Omega}_c\mathbf{S}_2=\left( {\begin{array}{*{20}{c}}
  s_1s_2\omega_{c;1}&0 \\ 
  0&\omega_{c;2}/s_1s_2 
\end{array}} \right) :=
\left( {{\begin{array}{*{20}{c}}
  \omega'_{c;1}&0 \\ 
  0&\omega'_{c;2}
\end{array}}} \right).
\end{align}
After dropping the primes for shortening and to keep the consistency of the notation, the expectation value of the corresponding test operator takes the form
\begin{align} \label{AExL}
  {\text{Tr}}( {{{\hat \varrho }_s}\hat {\mathcal L}} ) = {\omega _1}\left\langle {\left( {\hat q_1^2 + \hat p_1^2} \right)} \right\rangle  + {\omega _2}\left\langle {\left( {\hat q_2^2 + \hat p_2^2} \right)} \right\rangle \nonumber\\
  + {\omega _{c;1}} \left\langle {{\hat q}_1}{{\hat q}_2} \right\rangle + {\omega_{c;2}} \left\langle {{\hat p}_1}{{\hat p}_2} \right\rangle.
\end{align}

In general, the operator $\hat{\mathcal L}$ in Eq.~\eqref{ExL} can be written as the convex combinations of the form
\begin{equation}
 \hat{\mathcal L}=\alpha \hat{\mathcal L}_p{+}(1-\alpha)\hat{\mathcal L}_n,
\end{equation}
of two maximally correlated operators,
\begin{align}
 &\hat{\mathcal L}_p = {\omega _1}\left( {\hat q_1^2 {+} \hat p_1^2} \right) {+} {\omega _2}\left( {\hat q_2^2 {+} \hat p_2^2} \right) {+} {\omega _c}\left( {{{\hat q}_1}{{\hat q}_2} {+} {{\hat p}_1}{{\hat p}_2}} \right), \label{Lp}\\
 &\hat{\mathcal L}_n = {\omega _1}\left( {\hat q_1^2 {+} \hat p_1^2} \right) {+} {\omega _2}\left( {\hat q_2^2 {+} \hat p_2^2} \right) {+} {\omega_{c^\prime}}\left( {{{\hat q}_1}{{\hat q}_2} {-} {{\hat p}_1}{{\hat p}_2}} \right) \label{Ln},
\end{align}
where $0\leq \alpha \leq1$, $\omega_c=(\omega_{c;1}+\omega_{c;2})/\alpha$ and $\omega_{c^\prime}=(\omega_{c;1}-\omega_{c;2})/(1-\alpha)$.
Semiboundedness of $\hat{\mathcal L}_{p,n}$ then implies that $\omega_c^2{<}4\omega_1\omega_2$ and $\omega^2_{c^\prime}{<}4\omega_1\omega_2$.
In Appendix~\ref{AII} we have solved the SN eigenvalue problem for the operators in Eqs.~\eqref{Lp} and~\eqref{Ln}, which are semibounded from below.
Accordingly, two particular boundary SN eigenvalues for $\hat{\mathcal L}_{p}$ are obtained as 
\begin{equation}
 g_{1,\min } {=} \omega_1 {+} \omega_2, \qquad g_{\infty,\min } {=} \sqrt{{{{\left( {{\omega _1} {-} {\omega _2}} \right)}^2} {+} \omega _c^2}}.
\end{equation}
In the same way, for $\hat{\mathcal L}_{n}$ we have 
\begin{equation}
 g_{1,\min }{=}  \omega_1 {+} \omega_2, \qquad g_{\infty,\min }{=} \sqrt{{{{\left( {{\omega _1} {+}{\omega _2}} \right)}^2} {-} \omega^2_{c^{\prime}}}}.
\end{equation}
Now, given any SN $r$, $g_{r,\min }$ is the lowest expectation value, $\text{Tr}(\hat\varrho_s \hat{\mathcal L}_{p,n})$, which a state $\hat\varrho_s \in \mathcal{S}_r$ can attain.
Thus, if for the state under study the expectation value drops below $g_{r,\min }$, then its SN exceeds the value $r$.
Note that this method not only identifies entanglement, but also yields an entanglement measure in the form of a lower bound on the SN of the state.
The particular boundaries $g_{r,\min}$ are explicitly given in Appendix~\ref{AII} in forms of minors.

It is noteworthy that the action of the partial transposition map ${{\mathbf\Lambda} _j}\hat \varrho  = {{\hat \varrho }^{{T_j}}}$, with ${\mathbf\Lambda}_j{=}\text{diag}(1,\mp1,1,\pm1)$ ($j{=}1,2$), is equivalent to a mirror reflection of the $j$th subsystem in the phase space, or just the local time reversal of the $j$th subsystem~\cite{Duan,Simon2}. Applying ${\mathbf\Lambda}_j$ ($j{=}1,2$) to $\hat{\mathcal L}_{p,n}$ brings about a sign change of the determinant of $\mathbf{\Omega}_c$.
The other way around, under partial transposition (PT) transformation $\hat{\mathcal L}_p$ converts into $\hat{\mathcal L}_n$ and vice versa.

\section{Application}
\label{Exam}

In the following section, after the general argument of how to apply the witnesses based in Eqs.~\eqref{Lp} and~\eqref{Ln} to experimentally measured data, we examine examples of mixed Gaussian and non-Gaussian states.
To quantify the entanglement of a bipartite quantum state within an experiment, we follow these steps:
\begin{enumerate}[(i)]
 \item Measure the CM of the state using standard methods, e.g., homodyning. This is simply to measure the auto- or cross-correlations of the moments or quadratures up to second order.
 \item Transform the CM into its standard form, defined in Eq.~\eqref{StForm}.
 \item Choose the appropriate test operator between $\hat{\mathcal L}_{p}$ and $\hat{\mathcal L}_{n}$:
If $\det({\mathbf{V}_c}) {>} 0$, $\hat{\mathcal L}_{p}$ and if $\det({\mathbf{V}_c}) {<} 0$, $\hat{\mathcal L}_{n}$ should be choosen.
If $\det({\mathbf{V}_c}) {=} 0$ and ${\mathbf{V}_c} \ne 0$ one should examine both of the test operators $\hat{\mathcal L}_{p}$ and $\hat{\mathcal L}_{n}$.
 \item~\label{Stepiv} Construct the expectation value of the test operator 
\begin{equation}
  \text{Tr}(\hat\varrho_{\rm s} \hat{\mathcal L}_{p,n}) {=} 2(v_1\omega_1 {+} v_2\omega_2) {+} \omega_{c,c^\prime}(|v_{c,1}| {+} |v_{c,2}|).
\end{equation}
 \item Optimize the parameters of the test operator, namely $\omega _1$, $\omega _2$, and $\omega _{c}$ or $\omega _{c^\prime}$.
\end{enumerate}

Let us point out that, if ${\mathbf{V}_c} {=} 0$ the method is not useful to be applied. This is because $\mathbf{V}$ is a \textit{bona fide} CM if and only if $v_j\geq1/2$ ($j=1,2$)~\cite{Simon1}.
Therefore, vanishing cross correlations give rise to an expectation value of the test operators $\text{Tr}(\hat\varrho_{\rm s} \hat{\mathcal L}_{p,n})\geq g_{1,\text{min}}$, cf. step~\eqref{Stepiv}. Hence, no SN${>}1$ can be identified when there is no cross correlations in the CM. In fact, we always choose the test operator which gives the largest contribution from cross correlations.

So far, we have seen that a second optimization of the test operators ${{\hat {\mathcal L}}_{p,n}}$ is needed.
That is, for detecting the SN with the best precision, one should choose the optimized values for the parameters $\omega_j,j{=}1,2,c,c^\prime$.
This can be done by defining the distance from the lowest bound of the operator $\hat{\mathcal L}_{p,n}$ as
\begin{align} \label{Del}
  \Delta_{p,n}  &:= g_{\infty,\min } - {\text{Tr}}({{\hat \varrho }_s}{{\hat {\mathcal L}}_{p,n}}) \nonumber \\
  &= \sqrt {{{\left( {{\omega _1} \mp {\omega _2}} \right)}^2} \pm \omega _{c,c^\prime}^2}  \nonumber \\
  & \qquad \qquad - 2\left( {{v_1}{\omega _1} + {v_2}{\omega _2} + |{v_c}|{\omega _{c,c^\prime}}} \right).
\end{align}
in which $|v_c|:=(|v_{c,1}|+|v_{c,2}|)/2$. The optimal values for the parameters $\omega_j,j=1,2,c,c^\prime$ are obtained at the minimum point of this function.
Accordingly, this point is given by the solution of the equation $\Delta_{p,n}=0$.
We assume $\omega_{c,c^\prime}$ as the free parameter for the optimization and we simply obtain
\begin{align}
\label{A1stwc}
  &{\omega _{c,c^\prime}} =({{4v_c^2 \mp 1}})^{-1} \Bigg[- 4|{v_c}|\left( {{v_1}{\omega _1} + {v_2}{\omega _2}} \right) \nonumber \\
  &\mp\sqrt { \pm 4{{\left( {{v_1}{\omega _1} + {v_2}{\omega _2}} \right)}^2} + \left( {4v_c^2 \mp 1} \right){{\left( {{\omega _1} \mp {\omega _2}} \right)}^2}}
  \Bigg],
\end{align}
However, this is not the only condition.
One should also note that semiboundedness of $\hat{\mathcal L}_{p,n}$ requires that
\begin{align}
\label{ASBCond}
  \omega_{c,c^\prime}^2 < 4\omega_1 \omega_2.
\end{align}
Some states may fail to satisfy both these conditions.
In such cases, we find the best choices by solving the equation $\partial \Delta /\partial {\omega _{c^\prime}} = 0$ to get
\begin{align}
  &{\omega _{c}} =  - \frac{{2|{v_c}\left( {{\omega _1} - {\omega _2}} \right)|}}{{\sqrt {1 - 4v_c^2} }},\qquad |v_c| < \frac{1}{2},\\
  &{\omega _{c^\prime}} =  - \frac{{2|{v_c}|\left( {{\omega _1} + {\omega _2}} \right)}}{{\sqrt {1 + 4v_c^2} }}. \label{A2ndwc}
\end{align}

\subsection{Example of a Gaussian state}

To examine our SN test operators, consider a two-mode squeezed thermal state 
\begin{equation}
 \hat \varrho  = {{\hat S}_{ab}}( {\gamma ,\phi } )( {{{\hat \varrho }_{{\bar{n}_1}}} \otimes {{\hat \varrho }_{{\bar{n}_2}}}} )\hat S_{ab}^\dag ( {\gamma ,\phi } ),
\end{equation}
in which ${{\hat S}_{ab}}( {\gamma ,\phi } ) = \exp \{ {\gamma ( {{e^{i\phi }}{\hat{a}^\dag }{\hat{b}^\dag } - {e^{ - i\phi }}\hat{a}\hat{b}} )} \}$ is the two mode squeezing operator and
\begin{equation}
  {{\hat \varrho }_{{\bar{n}_j}}} = \frac{1}{{{{\bar n}_j} + 1}}\sum\limits_{n_j = 0}^\infty  {{{\left( {\frac{{{{\bar n}_j}}}{{{{\bar n}_j} + 1}}} \right)}^{n_j}}\left| n_j \right\rangle \left\langle n_j \right|},
\end{equation}
for $j=1,2$ is the thermal state of each mode with the mean thermal photon number $\bar{n}_j$.
This state is a mixed Gaussian state.
Adding locally an equal amount of thermal noise $\bar m$ to each mode, we will get the following CM elements~\cite{Marian,Glauber,Vourdas,Hall,Musslimani}:
\begin{align}
  &{v_{1,2}} = ({{\bar n}_{1,2}}{+}\frac{1}{2}){\cosh ^2}\gamma {+} ({{\bar n}_{2,1}}{+}\frac{1}{2}){\sinh ^2}\gamma {+} \bar m,\\
  &{v_c} =-{v'_c}= ( {{{\bar n}_1} + {{\bar n}_2} + 1} )\sinh \gamma \cosh \gamma,
\end{align}
with $\det({\mathbf{V}_c})={-}v^2_c$.
This gives ${\text{Tr}}( {{{\hat \varrho }_{\rm s}}{{\hat {\mathcal L}}_p}} ) = 2( {{\omega _1}{v_1} {+} {\omega _2}{v_2}} )$ which cannot detect entanglement.
On the other hand, one has 
\begin{equation}
 {\text{Tr}}( {{{\hat \varrho }_{\rm s}}{{\hat {\mathcal L}}_n}} ) {=} 2( {{\omega _1}{v_1} + {\omega _2}{v_2} {+} {\omega _{c^\prime}}{v_c}} ).
\end{equation}
Now, a simple minimization of the function $\Delta_n$ of Eq.~\eqref{Del} gives the best choices of the parameters.
It is easy to check that, for this state, the region of the thermal noise in which both of Eqs.~\eqref{A1stwc} and~\eqref{ASBCond} hold cannot exist.
A real $\omega_{c^\prime}$ requires (assuming $\bar n_1=\bar n_2=\bar n$)
\begin{align}
\label{STScond}
  4{\left[ {\left( {\bar n + \frac{1}{2}} \right)\left( {\cosh 2\gamma} \right) + \bar m} \right]^2} \leqslant {\left( {2\bar n + 1} \right)^2}\sinh^2 2\gamma.
\end{align}
In the simplest case, when $\bar n = \bar m = 0$, one has ${\cosh}^2 2\gamma  > {\sinh}^2 2\gamma $ which is in contradiction with Eq.~\eqref{STScond}.
Then, for a squeezed thermal state the best choice of the parameter $\omega_{c^\prime}$ is always given by Eq.~\eqref{A2ndwc}.

\begin{figure}[h]
  \includegraphics[width=8cm]{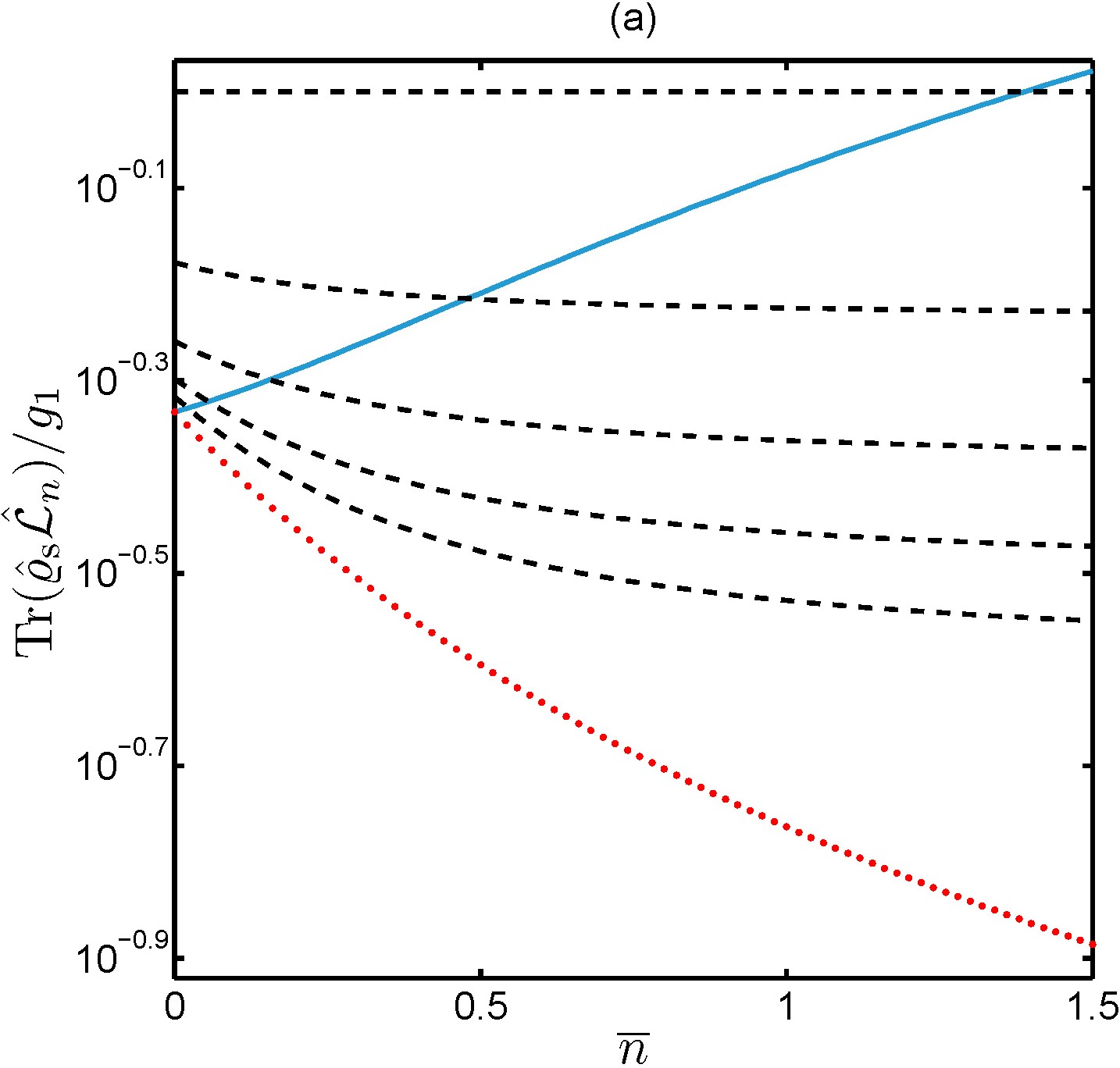}\\
  \vspace{0.3cm}
  \includegraphics[width=8cm]{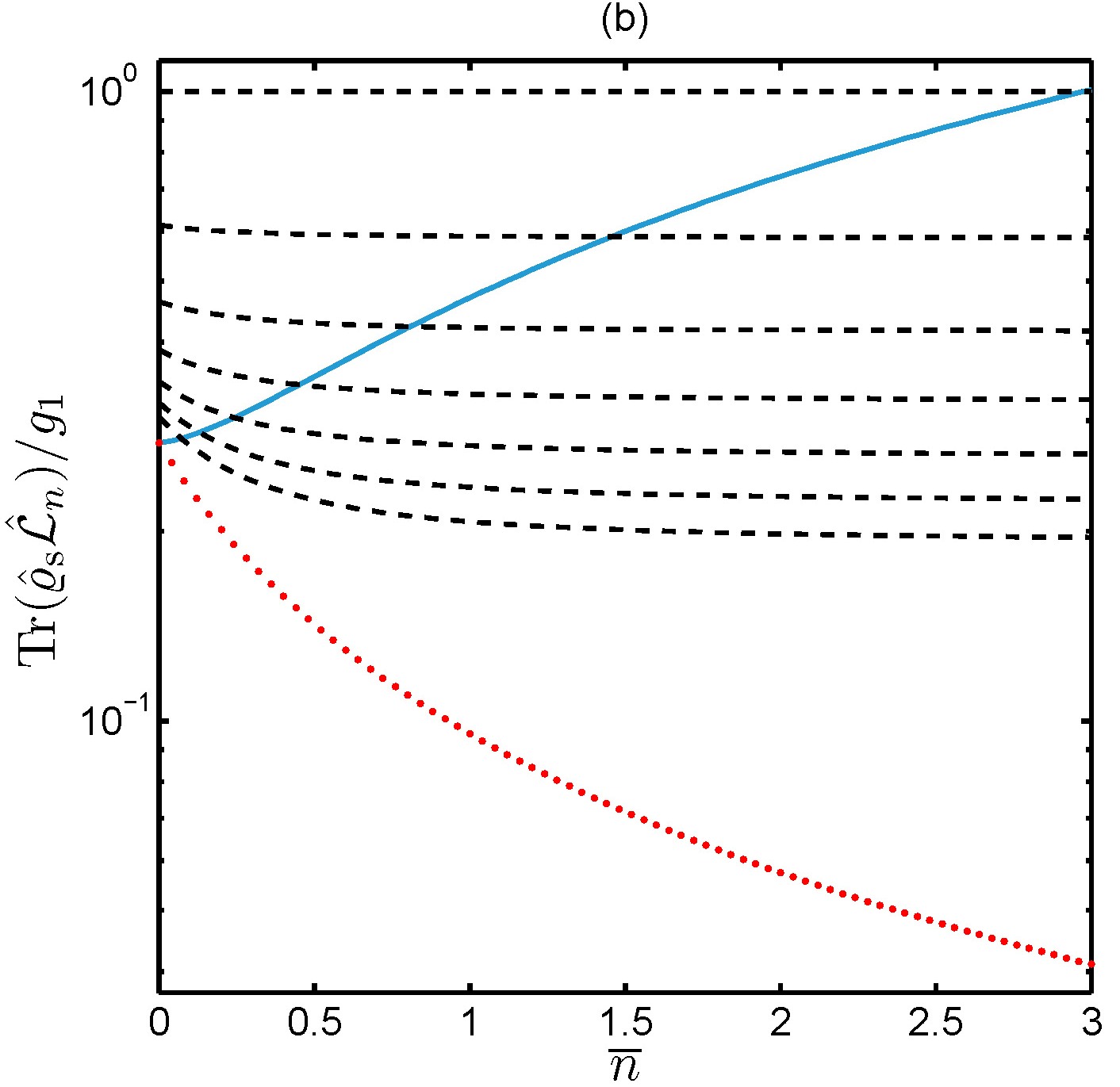}
  \caption{(Color online)
  Logarithmic sketches of the normalized expectation value of the test operator ${{\hat {\mathcal L}}_n}$ for squeezed thermal states ($\bar m = 0$) versus the mean global thermal noise (blue, full lines).
  The dashed lines represent the minimal SN-$r$ eigenvalues $g_r$.
  From top to bottom, the value of $r=1,2,\dots$ is increasing, the lowest (red, dotted) curves represent $g_{\infty}$. The
  squeezing parameters are
  (a) $\gamma{=}0.7$ and (b) $\gamma{=}0.98$.
  }\label{ThNoisen}
\end{figure}

In Fig.~\ref{ThNoisen} the changes of the normalized expectation value of ${{\hat {\mathcal L}}_n}$ for the squeezed thermal states versus the mean global thermal noise occupation $\bar n_1=\bar n_2=\bar n$ is depicted. 
The normalization is done with respect to $g_{1,\min}$.
The mean locally added noise $\bar m$ is taken to be zero and
the squeezing parameters are $\gamma{=}0.7$ and $0.98$, corresponding to squeezing powers of $9$ and $11.5$ dB, respectively.
For $\bar n=0$ one has an infinite value of the SN.
As global thermal noise increases, the SN decays rapidly.
However, one can see that the entanglement of the states (i.e., SN$>1$) can be identified up to $\bar n\cong1.4$ and $3$ for $\gamma=0.7$ and $0.98$, respectively.
Apparently, increasing the amount of squeezing yields an increase of the robustness of both the entanglement and its strength quantified by the SN against globally added noise.

\begin{figure}[h]
  \includegraphics[width=8cm]{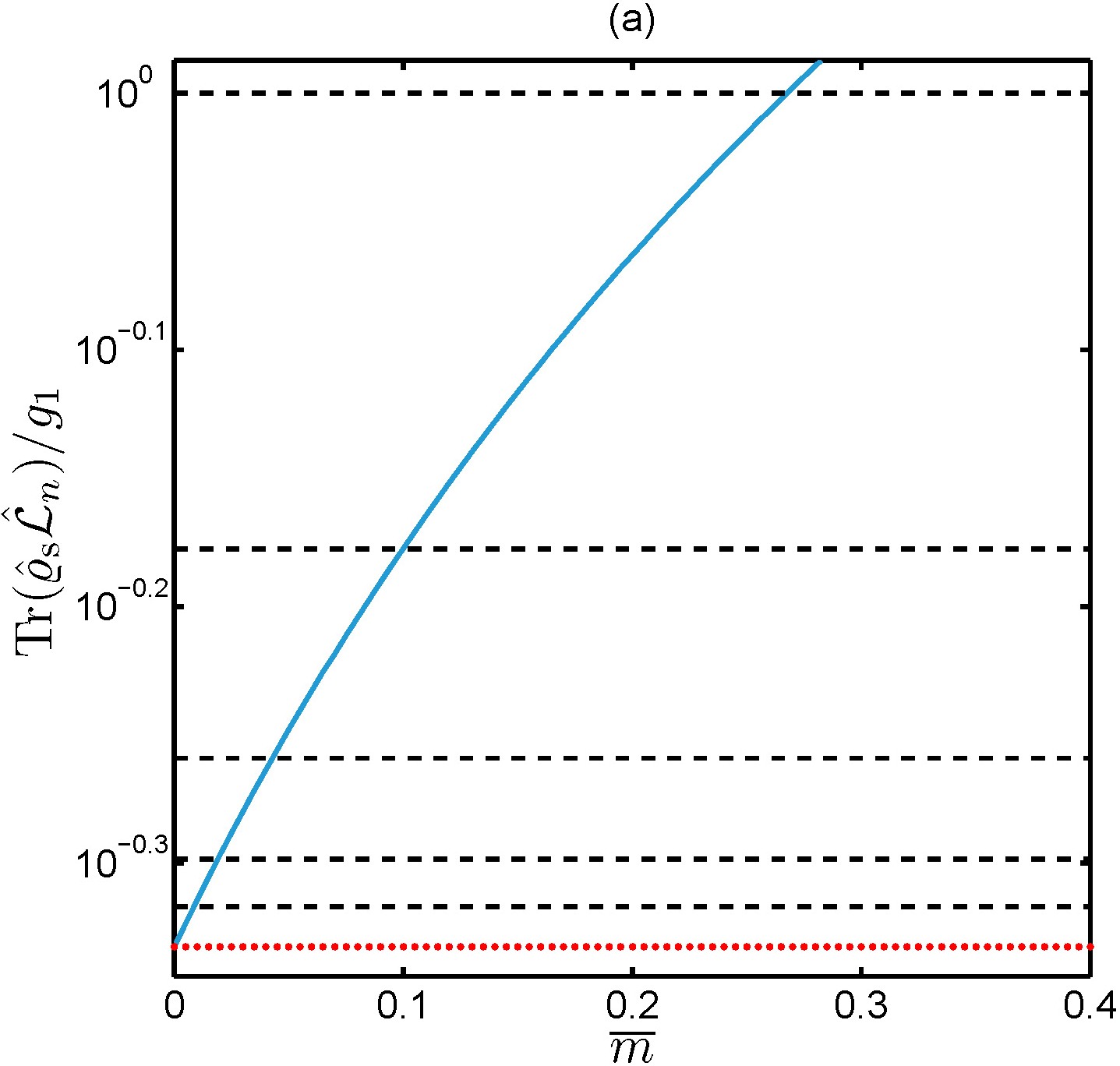}\\
  \vspace{0.3cm}
  \includegraphics[width=8cm]{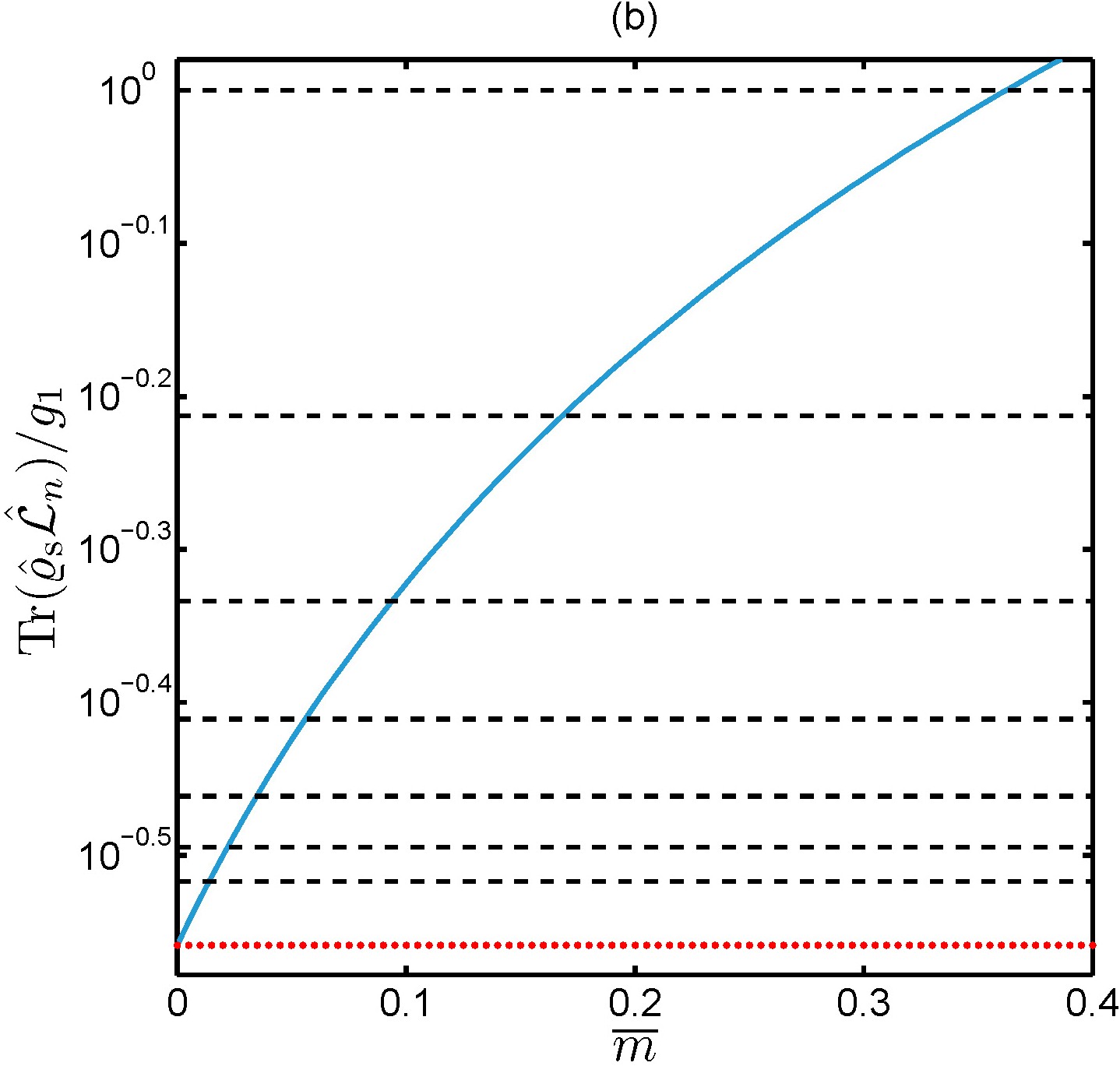}
  \caption{(Color online)
  Logarithmic sketches of the normalized expectation value of the test operator ${{\hat {\mathcal L}}_n}$ for squeezed thermal states ($\bar n_1{=} \bar n_2=0$) versus the mean thermal occupation of the locally added noise (blue, full lines).
  The dashed lines represent the minimal SN-$r$ eigenvalues $g_r$.
  From top to bottom, the value of $r=1,2,\dots$ is increasing, the lowest (red, dotted) curves represent $g_{\infty}$. The
  squeezing parameters are
  (a) $\gamma{=}0.7$ and (b) $\gamma{=}0.98$.
  }\label{ThNoisem}
\end{figure}

Figure~\ref{ThNoisem} shows the semilogarithmic sketch of the normalized expectation value of ${{\hat {\mathcal L}}_n}$ for squeezed thermal states versus the mean thermal occupation of the locally added noise $\bar m$.
In this case, the mean global thermal photon numbers are $\bar n_1{=} \bar n_2{=}0$.
For $\gamma{=}0.7$ and no added noise the SN is infinite.
However, the SN reduces rapidly with increasing thermal noise, so that only up to $\bar m{\cong}0.27$ a SN$>1$ can be detected.
With increasing the squeezing power, the robustness of the entanglement increases significantly.
For $\gamma{=}0.98$ the entanglement persists up to a mean thermal occupation of $\bar m{\cong}0.36$.
This shows once again that increasing the amount of squeezing yields an increase of the strength of entanglement quantified by the SN.

\begin{figure}[h]
  \includegraphics[width=8cm]{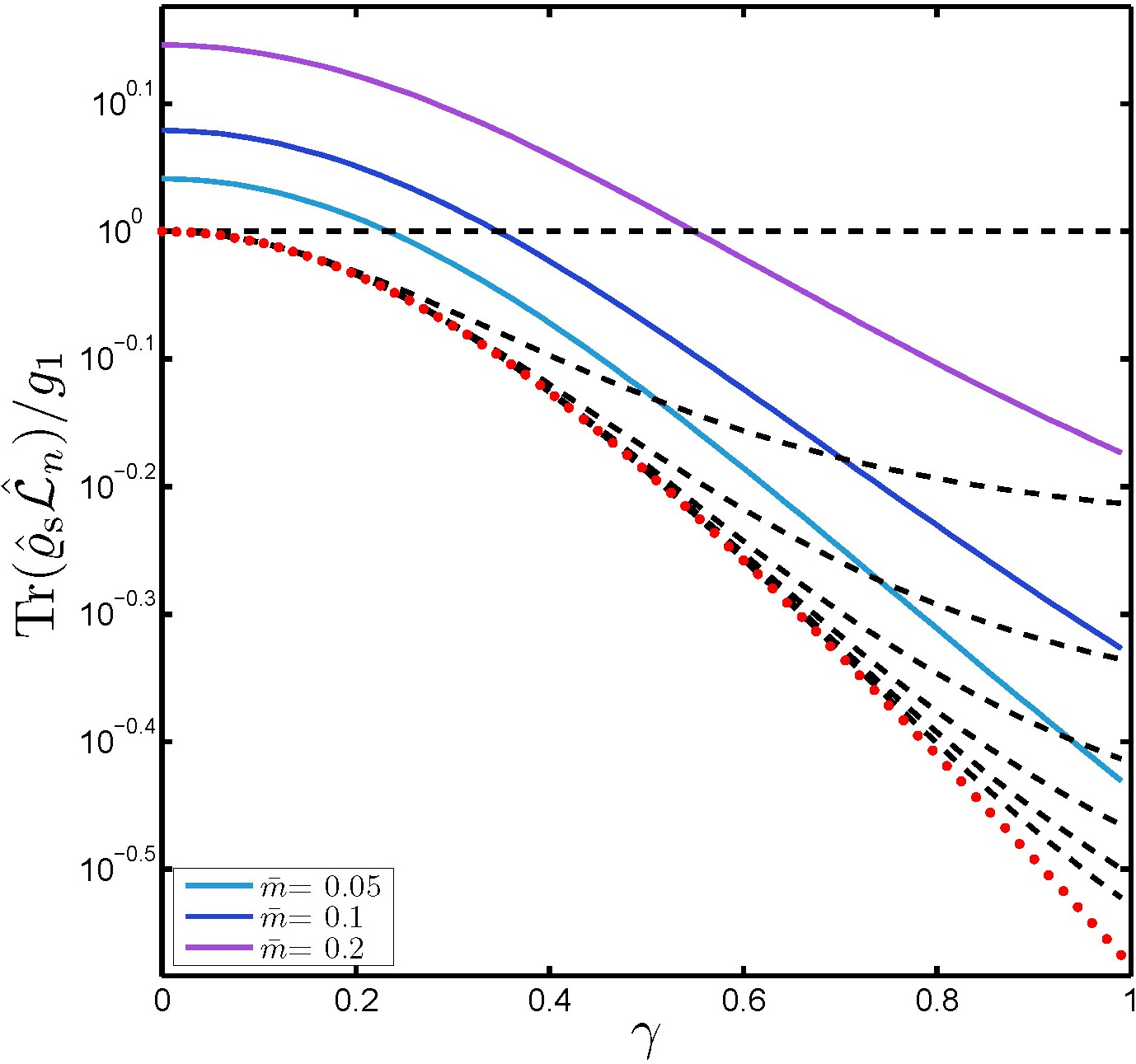}
  \caption{(Color online)
  The normalized SN levels and expectation values of ${{\hat {\mathcal L}}_n}$ for three squeezed thermal states versus squeezing parameter (full lines).
  The dashed lines represent the minimal SN-$r$ eigenvalues $g_r$.
  From top to bottom, the value of $r=1,2,\dots$ is increasing, the lowest (red, dotted) curve represents $g_{\infty}$. 
  Apparently, the entanglement is getting stronger with decreasing thermal noise.
  }\label{SqParameter}
\end{figure}

In Fig.~\ref{SqParameter} we demonstrate the change of normalized SN eigenvalues and normalized expectation values of the test operator with respect to the squeezing parameter $\gamma$ for three different mean thermal occupancies.
This variation is obviously due to the change of the optimized choice of parameters $\omega_j$ for $j{=}1,2,c^{\prime}$.
One can see that with decreasing thermal noise, larger SN bounds can be identified when the squeezing is increased.
For example, in the case of $\bar m{=}0.05$ in the interval $0{\leqslant} \gamma {\leqslant} 1$ the state crosses four SN levels and eventually reaches a SN of at least $r=5$. For $\bar m{=}0.1$ the state crosses only two SN level in the same interval.
This confirms the fact that the lower the thermal noise is, the larger is the amount of entanglement quantified by the SN, for a given squeezing strength.

\subsection{Application to non-Gaussian states}
As pointed out before, one can apply both tests ${{\hat {\mathcal L}}_{p}}$ and ${{\hat {\mathcal L}}_{n}}$ to certify a lower bound for the SN of non-Gaussian states. 
A non-Gaussian state possesses higher order moments which do not appear in the CM and thus needs higher order witnesses~\cite{Gomes,Miranowicz,Agarwal,Hillery}. 
Nevertheless, if the cross correlation elements of the CM exist, one can follow this method and quantify the entanglement.
We apply our method to a partially phase-randomized squeezed vacuum state as an example of a non-Gaussian state.
This state may be produced by phase randomization in one channel of a two-mode squeezed vacuum state~\cite{Kiesel,Mehmet,sp-vo-NJP}.
Such a state with a Gaussian phase-diffusion of variance $\sigma^2$ has a Fock basis representation of the form 
\begin{equation}
{{\hat \varrho }_\sigma } = (1 {-} {\epsilon ^2})\sum_{m,n} {{\epsilon ^{m {+} n}} e^{\{ { {-} {\sigma ^2}{{(m {-} n)}^2}/2} \}}\left| {m,m} \right\rangle \left\langle {n,n} \right|}, 
\end{equation}
with $0 {\leqslant} \epsilon  {<} 1$ being related to the squeezing parameter, $\epsilon=\tanh \gamma$.
In the limit of $\sigma \to \infty$, the resulting fully phase-randomized squeezed vacuum state is separable.

It is straightforward to calculate the standard form of the CM for ${{\hat \varrho }_\sigma }$ given by the parameters
\begin{align}
 &{v_{1,2}} {=} (1 {+} {\epsilon ^2}){/}[2(1 {-} {\epsilon ^2})],\\
 &{v_c} = {-} {v_c'} = 2\epsilon (1 {-} {\epsilon ^2})^{-1} e^{{-}\sigma ^2/2}.
\end{align}
Likewise, we should use ${{\hat {\mathcal L}}_n}$ as the appropriate test operator.
Then, one finds
\begin{equation}
 \text{Tr}(\hat\varrho_{\sigma ,\rm s} \hat{\mathcal L}_{n}){=}2 [ (\omega_1 {+} \omega_2)v_j{+}\omega_{c^\prime} v_c].
\end{equation}
For the best choice of the parameters $\omega_j$ ($j{=}1,2,c^{\prime}$), a minimization of the function $\Delta_n$ given in Eq.~\eqref{Del} is required.
In turn, if we examine Eq.~\eqref{A1stwc} for partially phase-randomized squeezed vacuum states, after a short calculation we find that this equation cannot hold.
That is, one needs $\exp \{- \sigma^2 \} >1$ for $\sigma \ne 0$ which, regarding $\sigma$ being a real parameter, is not possible.
So, as in the case of squeezed thermal states, one should use Eq.~\eqref{A2ndwc} to obtain the optimized value of the parameter $\omega_{c^\prime}$.

\begin{figure}[h]
  \includegraphics[width=8cm]{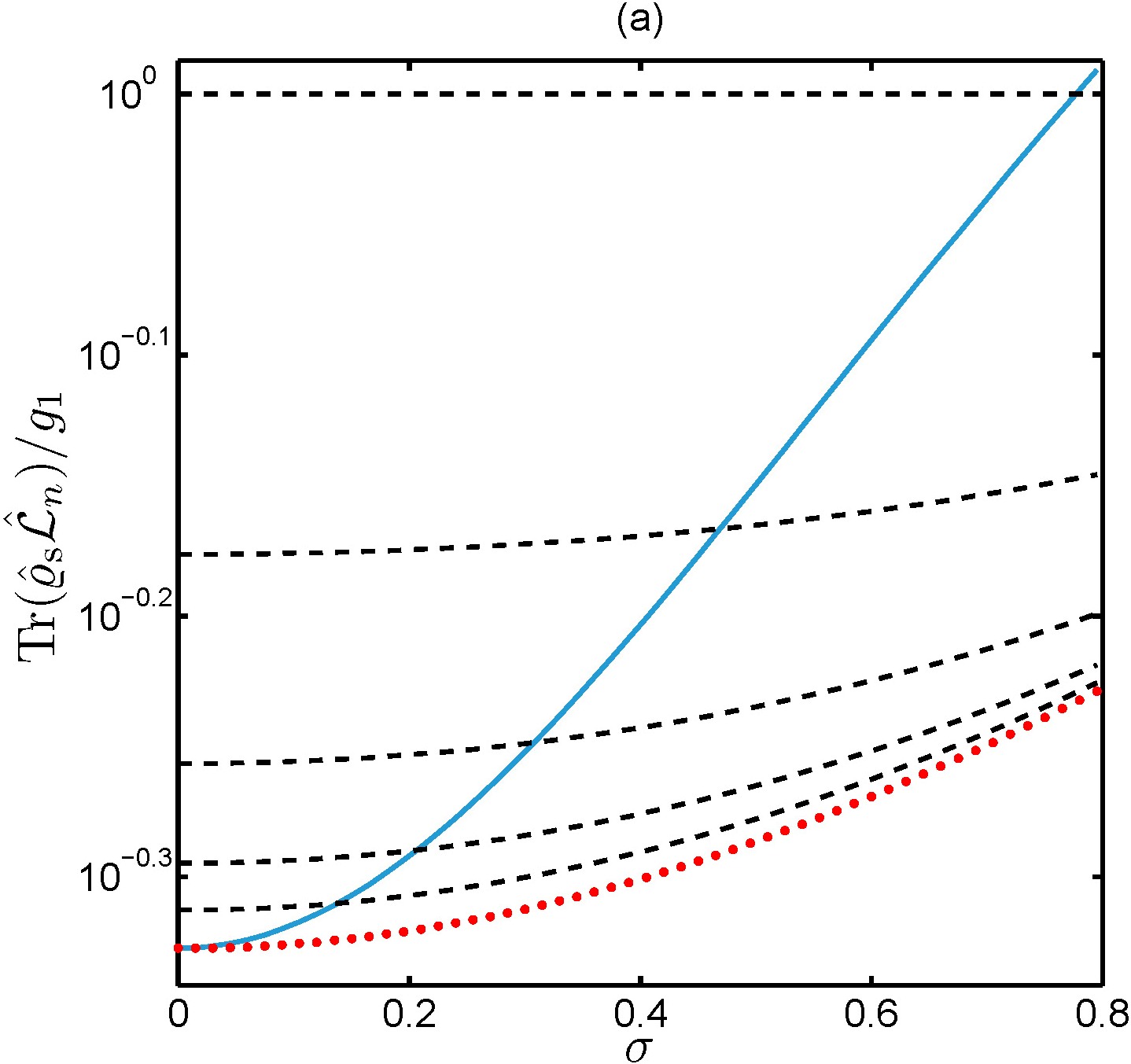}\\
  \vspace{0.3cm}
  \includegraphics[width=8cm]{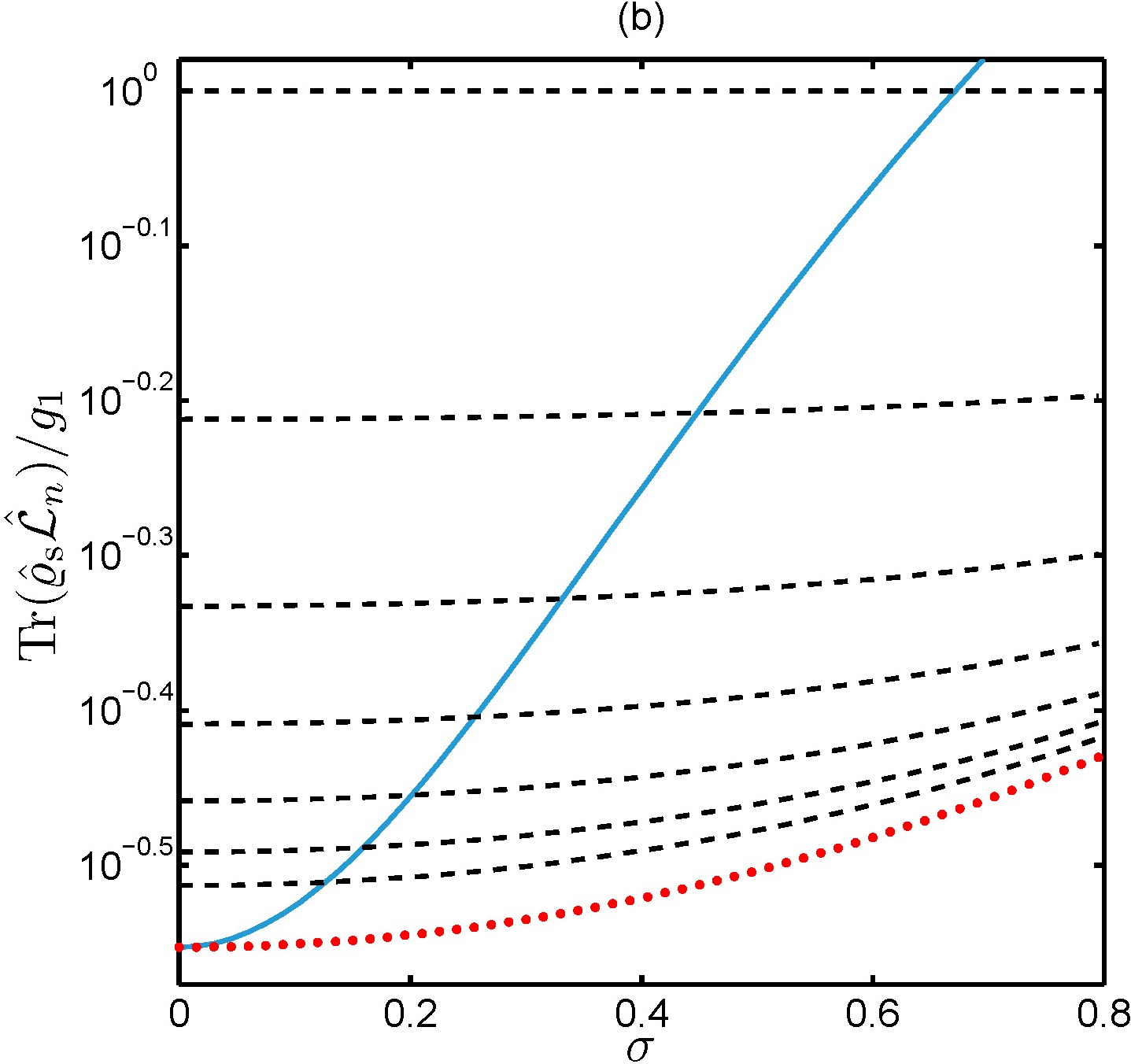}
  \caption{(Color online)
  Logarithmic sketches of the normalized expectation value of the test operator ${{\hat {\mathcal L}}_n}$ for a phase randomized, bipartite squeezed-state ${{\hat \varrho }_\sigma }$ versus phase-randomization parameter $\sigma$ (blue, full lines).
  The dashed lines represent the minimal SN-$r$ eigenvalues $g_r$.
  From top to bottom, the value of $r=1,2,\dots$ is increasing, the lowest (red, dotted) curves represent $g_{\infty}$. For the
  squeezing parameter of 
  (a) $\gamma{=}0.7$ and (b) $\gamma{=}0.98$ 
  entanglement can be detected up to a phase randomization of $\sigma {\cong}0.775$ and $\sigma{\cong}0.665$, respectively.
  }\label{PR}
\end{figure}

In Fig.~\ref{PR} we present the results for two values of the squeezing strength, $\gamma=0.7$ and $0.98$.
As discussed above, even though the state is non-Gaussian, one can detect the entanglement and quantify it using the SN, solely by measuring the CM of the state.
Dashed lines represent SN-$r$ eigenvalues.
When the expectation value is below $g_{r}$, the SN exceeds the corresponding $r$-value, $\text{SN}>r$.
For the squeezing parameter $\gamma=0.7$, we can detect entanglement up to a phase-randomization $\sigma$ of about $0.775$.
It is also possible to detect the entanglement up to $\sigma {\cong} 0.665$ at $\gamma=0.98$.
It is reasonable that with increasing strength of squeezing the fragility of the state against dephasing increases.
However, for stronger squeezing and small values of the phase randomization, $\sigma$, a larger SN can be identified.
Note that the entanglement of the state under study could be identified by entanglement quasiprobabilities even for $\sigma=5$, see \cite{sp-vo-NJP}.
However, the Gaussian SN witnesses are a powerful tool to guaranty a certain Schmidt number.
The method is relatively simple, since it only requires to measure the covariance matrix elements of the state to be analyzed.
It also gives an insight into the amount of entanglement for non-Gaussian quantum states.

\section{Conclusions}
\label{Summ}

In conclusion, we have characterized the general class of bipartite Gaussian Schmidt-number witnesses.
We have shown that there exist two general forms of such operators.
We draw the close connection between Gaussian Schmidt-number tests and the covariance matrix of quantum states, which is experimentally accessible.
Furthermore, we studied the application of our method to observable CMs.
In this sense, Gaussian Schmidt-number witnesses are a simple operational tool for detecting and quantifying entanglement of bipartite continuous variable systems.
This fills the gap between theoretical and experimental methods for quantification of entanglement.

The power of the method is demonstrated by application to two-mode squeezed thermal states as an example of bipartite Gaussian states.
This gives insight into the needed squeezing and boundaries to the noise level for which a certain amount of entanglement can be guaranteed.
The Gaussian witness operators are also useful for entanglement quantification of some particular non-Gaussian states, as it is demonstrated for phase-randomized squeezed vacuum states.
Our method identifies a lower bound of the Schmidt number for any quantum state, but not necessarily the greatest one for non-Gaussian states.
The advantage of our method is the possibility of entanglement quantification by simply measuring covariance matrices, which is practically simple.

\section*{Acknowledgments}
The authors gratefully acknowledge fruitful discussions with E. Agudelo, P. Gr\"unwald and D. Vasylyev.
This work was supported by the Deutsche Forschungsgemeinschaft through SFB~652.

\appendix
\begin{widetext}

%
\section{Solution of the SN-$r$ eigenvalue problem}\label{AII}

Here, we solve the SN eigenvalue problem~\cite{Sperling1} for the two forms of the Gaussian test operators, c.f. Eqs.~\eqref{Lp} and~\eqref{Ln}.
We assume that the solutions of Eq.~\eqref{SNEE1} are orthogonal and the solutions of Eq.~\eqref{SNEE2} are orthonormal.
These are well-justified assumptions due to the existence of Schmidt decomposition for the final solution.
For $\hat{\mathcal L}_p$, after representing the test operator in terms of boson operators, $\hat a\equiv (\hat q_1+i\hat p_1)/\sqrt{2}$ and $\hat b\equiv (\hat q_2+i\hat p_2)/\sqrt{2}$, one obtains

\begin{align}
&\left\{ {2{\omega _1}\mathbf{N}_1^{(r)} + 2{\omega _2}{\hat b^\dag }\hat b + {\omega _c}\left( {{\mathbf{A}^{(r)}}{\hat b^\dag } + {\mathbf{A}^{(r)\dag }}\hat b} \right) + {\omega _1} + {\omega _2}} \right\}| {\vec{\zeta _r}} \rangle  = {g_{r}}| {\vec{\zeta _r}} \rangle, \label{ASNEEM1}\\
&\left\{ {2{\omega _1}\mathbf{N}_2^{(r)} + 2{\omega _2}{\mathbf{O}_2^{(r)}}{\hat a^\dag }\hat a + {\omega _c}\left( {{\mathbf{B}^{(r)}}{\hat a^\dag } + {\mathbf{B}^{(r)\dag }}\hat a} \right) + ({\omega _1} + {\omega _2}}){\mathbf{O}_2^{(r)}} \right\}| {\vec{\xi _r}} \rangle  = {g_{r}}{\mathbf{O}_2^{(r)}}| {\vec{\xi _r}} \rangle,
\end{align}
in which
\begin{align}
  &\mathbf{N}_{1}^{(r)}=\left( {\begin{array}{*{20}{c}}
  \left\langle \xi_1 \right| \hat a^\dag \hat a \left| \xi_1 \right\rangle & \cdots & \left\langle \xi_1 \right| \hat a^\dag \hat a \left| \xi_r \right\rangle \\ 
  \vdots & \ddots & \vdots \\
  \left\langle \xi_r \right| \hat a^\dag \hat a \left| \xi_1 \right\rangle & \cdots & \left\langle \xi_r \right| \hat a^\dag \hat a \left| \xi_r \right\rangle
\end{array}} \right), \nonumber \\
  &\mathbf{N}_{2}^{(r)}= \left( {\begin{array}{*{20}{c}}
  \left\langle \zeta_1 \right| \hat b^\dag \hat b \left| \zeta_1 \right\rangle & \cdots & \left\langle \zeta_1 \right| \hat b^\dag \hat b \left| \zeta_r \right\rangle \\ 
  \vdots & \ddots & \vdots \\
  \left\langle \zeta_r \right| \hat b^\dag \hat b \left| \zeta_1 \right\rangle & \cdots & \left\langle \zeta_r \right| \hat b^\dag \hat b \left| \zeta_r \right\rangle
\end{array}} \right), \nonumber \\
  &\mathbf{A}^{(r)}=\left( {\begin{array}{*{20}{c}}
  \left\langle \xi_1 \right| \hat a \left| \xi_1 \right\rangle & \cdots & \left\langle \xi_1 \right| \hat a \left| \xi_r \right\rangle \\ 
  \vdots & \ddots & \vdots \\
  \left\langle \xi_r \right| \hat a \left| \xi_1 \right\rangle & \cdots & \left\langle \xi_r \right| \hat a \left| \xi_r \right\rangle
\end{array}} \right), \nonumber \\
  &\mathbf{B}^{(r)}= \left( {\begin{array}{*{20}{c}}
  \left\langle \zeta_1 \right| \hat b \left| \zeta_1 \right\rangle & \cdots & \left\langle \zeta_1 \right| \hat b \left| \zeta_r \right\rangle \\ 
  \vdots & \ddots & \vdots \\
  \left\langle \zeta_r \right| \hat b \left| \zeta_1 \right\rangle & \cdots & \left\langle \zeta_r \right| \hat b \left| \zeta_r \right\rangle
\end{array}} \right), \nonumber \\
  &\mathbf{O}_{2}^{(r)}= \left( {\begin{array}{*{20}{c}}
  \left\langle \zeta_1 \right. \left| \zeta_1 \right\rangle & \cdots & \left\langle \zeta_1 \right. \left| \zeta_r \right\rangle \\ 
  \vdots & \ddots & \vdots \\
  \left\langle \zeta_r \right. \left| \zeta_1 \right\rangle & \cdots & \left\langle \zeta_r \right. \left| \zeta_r \right\rangle
\end{array}} \right),
\end{align}  
We can transform the matrix ${\mathbf{A}^{(r)}}$ of Eq.~\eqref{ASNEEM1} in an upper triangular form, and consequently ${\mathbf{A}^{(r)\dag }}$ in a lower triangular form by using a unitary transformation of the basis vectors of the first system, because this transformation does not change the SN eigenvalue equation or the SN eigenvalues. Then, the solution leads to SN-$r$ eigenvectors as $\left| {{\varphi _r}} \right\rangle  = \sum\limits_{n = 1}^r {\varphi _n^{(r)}\left| n{-}1 \right\rangle \left| {r {-} n} \right\rangle } $. The key point to obtain the solution is that Eq.~\eqref{ASNEEM1} contains the operator $\hat b^\dag \hat b$ as its diagonal elements, so that the only possibility for $|\zeta_i \rangle$ to be its eigenvector is a Fock state. Moreover, the appearance of $\hat b^\dag$ ($b$) on the upper (lower) triangle imposes that $\left\langle \xi_i \right| \hat a \left| \xi_ j\right\rangle{=}0$ when $ i {\neq} j+1 $ ($\left\langle \xi_i \right| \hat a^\dag \left| \xi_ j\right\rangle{=}0$ when $ i {\
neq}
 j-1 $). Then, the 
matrix form of this equation is obtained as
\begin{align}
 \left( {\begin{array}{*{20}{c}}
  {2{\omega _2}{\hat b^\dag }\hat b}&{{\omega _c}{\hat b^\dag }}&{}&{}&{}&{} \\ 
  {{\omega _c}\hat b}& \ddots &{}&{}&{}&{} \\ 
  {}&{}&2[{(i {-} 1){\omega _1} {+} {\omega _2}{\hat b^\dag }\hat b}]&{\sqrt i {\omega _c}{\hat b^\dag }}&{}&{} \\ 
  {}&{}&{\sqrt i {\omega _c}\hat b}&2[{i{\omega _1} {+} {\omega _2}{\hat b^\dag }\hat b}]&{}&{} \\ 
  {}&{}&{}&{}& \ddots &{\sqrt {r {-} 1} {\omega _c}{\hat b^\dag }} \\ 
  {}&{}&{}&{}&{\sqrt {r {-} 1} {\omega _c}\hat b}&2[{(r {-} 1){\omega _1} {+} {\omega _2}{\hat b^\dag }\hat b}] 
\end{array}} \right)
\left( {\begin{array}{*{20}{c}}
  {\varphi _1^{(r)}\left| r{-}1 \right\rangle } \\ 
  {\vdots} \\ 
  {\varphi _i^{(r)}\left| r{-}i \right\rangle } \\ 
  {\varphi _{i{-}1}^{(r)}\left| r{-}i{-}1 \right\rangle }  \\ 
  {\vdots} \\ 
  {\varphi _r^{(r)}\left| 0 \right\rangle } 
\end{array}} \right) \nonumber \\ = g'
\left( {\begin{array}{*{20}{c}}
  {\varphi _1^{(r)}\left| r{-}1 \right\rangle } \\ 
  {\vdots} \\ 
  {\varphi _i^{(r)}\left| r{-}i \right\rangle } \\ 
  {\varphi _{i{-}1}^{(r)}\left| r{-}i{-}1 \right\rangle }  \\ 
  {\vdots} \\ 
  {\varphi _r^{(r)}\left| 0 \right\rangle } 
\end{array}} \right),
\end{align}
where $g'_r=g_r - \omega_1-\omega_2$. Eigenvalues are then given by the following $r$th order determinant:
\begin{align}
\label{ADetp}
\left| {\begin{array}{*{20}{c}}
  {2(r {-} 1){\omega _2} {-} g'_r}& {\omega_c}&{}&{}&{}&{} \\ 
  \omega_c& \ddots &{}&{}&{}&{} \\ 
  {}& {} &{2[(i {-} 1){\omega _1} + (r - i){\omega _2}] - g'_r}&{\sqrt {(r {-} i )i}{\omega _c} }&{}&{} \\ 
  {}&{}&{\sqrt {(r {-} i )i}{\omega _c} }&{2[i{\omega _1} {+} (r {-} i {-}1){\omega _2}] {-} g'_r}& {} &{} \\ 
  {}&{}&{}&{}& \ddots &\sqrt {r {-} 1} {\omega _c} \\ 
  {}&{}&{}&{}&\sqrt {r {-} 1} {\omega _c}&{2(r {-} 1){\omega _1} {-} g'_r} 
\end{array}} \right|=0.
\end{align}
Note that, in principle, the SN-$r$ eigenvector could be constructed from any subset of Fock states $\{|n\rangle,\dots,|n{+}r{-}1\rangle\}~(n \in \{0,1,2,\dots\})$. However, since we are looking for the minimum possible eigenvalue, we have chosen the set of Fock states to begin from zero $(n=0)$.

Concerning the mathematical treatment to solve SN-$r$ eigenvalue equations, there is no restriction on the parameters. However, since we are interested in the bounded Hermitian operators for witnessing, we assume that $\omega_j>0$ $(j{=}1,2)$. Then, the resulting class of operators will be semibounded from below.
Two limiting cases for optimal SN-$r$ eigenvalues are then
\begin{equation}
  g_{1,\min } = \omega_1 + \omega_2, \quad
  g_{\infty,\min } =  \sqrt {{{\left( {{\omega _1} - {\omega _2}} \right)}^2} + \omega _c^2}.
\end{equation}
Repeating the same procedure for $\hat{\mathcal L}_n$ results in the set of SN-$r$ eigenvectors of the form $\left| {{\psi _r}} \right\rangle  = \sum\limits_{n = 1}^r {\psi _n^{(r)}\left| n-1 \right\rangle \left| n-1 \right\rangle } $. The matrix equation is given by
\begin{align}
 \left( {\begin{array}{*{20}{c}}
  {2{\omega _2}{\hat b^\dag }\hat b}&{{\omega _{c^\prime}}{\hat b }}&{}&{}&{}&{} \\ 
  {{\omega _{c^\prime}}\hat b^\dag}& \ddots &{}&{}&{}&{} \\ 
  {}&{}&2[{(i {-} 1){\omega _1} {+} {\omega _2}{\hat b^\dag }\hat b}]&{\sqrt i {\omega _{c^\prime}}{\hat b }}&{}&{} \\ 
  {}&{}&{\sqrt i {\omega _{c^\prime}}\hat b^\dag}&2[{i{\omega _1} {+} {\omega _2}{\hat b^\dag }\hat b}]&{}&{} \\ 
  {}&{}&{}&{}& \ddots &{\sqrt {r {-} 1} {\omega _{c^\prime}}{\hat b }} \\ 
  {}&{}&{}&{}&{\sqrt {r {-} 1} {\omega _{c^\prime}}\hat b^\dag}&2[{(r {-} 1){\omega _1} {+} {\omega _2}{\hat b^\dag }\hat b}] 
\end{array}} \right)
\left( {\begin{array}{*{20}{c}}
  {\psi _1^{(r)}\left| 0 \right\rangle } \\ 
  {\vdots} \\ 
  {\psi _i^{(r)}\left| i{-}1 \right\rangle } \\ 
  {\psi _{i{-}1}^{(r)}\left| i \right\rangle }  \\ 
  {\vdots} \\ 
  {\psi _r^{(r)}\left| r{-}1 \right\rangle } 
\end{array}} \right) \nonumber \\ = g'
\left( {\begin{array}{*{20}{c}}
  {\psi _1^{(r)}\left| 0 \right\rangle } \\ 
  {\vdots} \\ 
  {\psi _i^{(r)}\left| i{-}1 \right\rangle } \\ 
  {\psi _{i{+}1}^{(r)}\left| i \right\rangle }  \\ 
  {\vdots} \\ 
  {\psi _r^{(r)}\left| r{-}1 \right\rangle } 
\end{array}} \right).
\end{align}
in which $g'_r=g_r - \omega_1-\omega_2$. Therefore, the eigenvalues are given through the $r$th order determinant
\begin{align}
\label{Detn}
\left| {\begin{array}{*{20}{c}}
  { - g'_r}&\omega_{c^\prime}&{}&{}&{}&{} \\ 
  \omega_{c^\prime}& \ddots &{}&{}&{}&{} \\ 
  {}& {} &{2(i {-} 1)({\omega _1} + {\omega _2}) {-} g'_r}&{i{\omega _{c^\prime}}}&{}&{} \\ 
  {}&{}&{i{\omega _{c^\prime}}}&{2i({\omega _1} {+} {\omega _2}) {-} g'_r}&{}&{} \\ 
  {}&{}&{}&{}& \ddots &{(r {-} 1){\omega _{c^\prime}}} \\ 
  {}&{}&{}&{}&{(r {-} 1){\omega _{c^\prime}}}&{2(r {-} 1)({\omega _1} {+} {\omega _2}) {-} g'_r} 
\end{array}} \right| = 0.
\end{align}
In this case, the limiting optimal eigenvalues are
\begin{align}
  g_{1,\min } =  \omega_1 + \omega_2, \quad
  g_{\infty,\min } = \sqrt {{{\left( {{\omega _1} +{\omega _2}} \right)}^2} - \omega _{c^\prime}^2}.
\end{align}
Note that for $r=\infty$, the corresponding minimum SN eigenvalue is just the ground state of the test operator, cf. Ref.~\cite{Lu}.
\end{widetext}

\end{document}